\documentclass[letterpaper, 10 pt, conference]{ieeeconf}  

\IEEEoverridecommandlockouts                              
\usepackage{verbatim}
\usepackage{graphics} 
\usepackage{amsmath,amssymb,stmaryrd,mathtools, amsfonts}
\usepackage{multirow}
\usepackage{hhline}
\usepackage{algorithm}
\usepackage[noend]{algpseudocode}
\PassOptionsToPackage{hyphens}{url}\usepackage{hyperref} 
\usepackage{dsfont}
\usepackage{algorithmicx}
\usepackage{multirow}
\usepackage[utf8]{inputenc}
\usepackage{import}
\usepackage{graphicx, graphics}
\usepackage{subcaption}
\usepackage{array}
\usepackage{color,xcolor}
\usepackage{soul}
\usepackage{booktabs}
\let\labelindent\relax
\usepackage{enumitem}
\definecolor{v0}{rgb}{1, 0.5, 0} 
\definecolor{v1}{rgb}{0, 1, 0.83} 
\definecolor{v2}{rgb}{0.25, 0.25, 0.25} 
\definecolor{v3}{rgb}{1, 0 ,0.58 } 

\usepackage{siunitx} 

\usepackage{svg}
\usepackage[noadjust]{cite}
\title{\LARGE \bf
Increasing Electric Vehicles Utilization in Transit Fleets using Learning, Predictions, Optimization, and Automation}

\author{Jacopo Guanetti\textsuperscript{2}, Yeojun Kim\textsuperscript{2}, Xu Shen\textsuperscript{1}, Joel Donham\textsuperscript{2}, Santosh Alexander\textsuperscript{2}, \\ Bruce Wootton\textsuperscript{2}, Francesco Borrelli\textsuperscript{1}
\thanks{\textsuperscript{1} Department of Mechanical Engineering, University of California, Berkeley, USA \texttt{\{xu\_shen, fborrelli\}@berkeley.edu}}
\thanks{\textsuperscript{2} WideSense, Inc. Berkeley, CA, USA.}
\thanks{This material is based upon work partially supported by 
(1) ARPA-E NEXTCAR Program - DE-AR0000791,
(2) the National Science Foundation under Grant 2019458, 
(3) the Department of Energy, Office of Science under Grant DE-SC0020894.
Any opinions, findings, conclusions or recommendations presented in this material are only those of the authors and do not necessarily reflect the views of the ARPA-E, NSF or DOE.}
\thanks{Francesco Borrelli has financial interests in WideSense, Inc.}
}

\begin{document}

\maketitle
\thispagestyle{empty}
\pagestyle{empty}


\begin{abstract}
This work presents a novel hierarchical approach to increase Battery Electric Buses (BEBs) utilization in transit fleets.
The proposed approach relies on three key components.
A learning-based \emph{BEB digital twin} cloud platform is used to accurately predict BEB charge consumption on a per vehicle, per driver, and per route basis, and accurately predict the time-to-charge BEB batteries to any level. 
These predictions are then used by a \emph{Predictive Block Assignment} module to maximize the BEB fleet utilization.
This module computes the optimal BEB daily assignment and charge management strategy. 
A \emph{Depot Parking and Charging Queue Management} module is used to autonomously park and charge the vehicles based on their charging demands.
The paper discusses the technical approach and benefits of each level in the architecture and concludes with a realistic simulations study.
The study shows that if our approach is employed BEB fleet utilization can increase by a 50\% compared to state-of-the-art methods.

\end{abstract}

\section{Introduction}
Transportation has become the largest polluter of Greenhouse Gas Emissions (GHG) in the US since 2016. 
Electrification is the primary solution to the decarbonization of transportation.
Governments around the world are encouraging electric mobility by enacting regulations and providing funds to road transit agencies to transition their fleets to zero-emission vehicles (ZEV). 

Transit agencies started deploying relatively small numbers of ZEVs in their fleets in the last few years; as they transition to ZEVs a growing percentage of their fleets, they are increasingly focusing on strategies to efficiently and cost-effectively meet service requirements designed around conventional vehicles, using zero-emission technology.
In transit agencies with a fixed schedule, a \emph{trip} covers a certain sequence of stops at pre-defined times, along a specific route.
Trips that are adjacent in space and time are organized into \emph{blocks}, units of work to be performed by a single bus.
Normally, blocks are designed with diesel buses in mind, and they may be unfeasible for the driving range Battery Electric Buses (BEBs), particularly in challenging conditions such as winter operation in cold regions.
Diesel buses that have completed a block can be quickly refueled and used on subsequent blocks.
BEBs take longer to recharge than diesel buses to refuel, and the charging time also depends on factors including the battery State of Charge (SOC), on the ambient and battery temperature; the charging cost also depends non-linearly on the power demand.
In general, ZEV includes BEB as well as other clean fuel vehicles such as hydrogen vehicles; this paper is focused on BEBs and the terms ZEV and BEB are used interchangeably.

Uncertainty and variability of driving range and charging time are major hurdles for transit agencies in planning for and operating BEBs.
Driving range can vary by a factor of 5 depending on a host of conditions, including temperature and weather, traffic, route profile, driving behavior, occupancy, and the configuration and condition of the vehicle and its components (such as battery aging).
Transit agencies need reliable service and are currently unable to accurately predict the energy and charging time required for each piece of work; instead, they rely on simplistic plans and nominal estimates of miles-per-gallon-equivalent (MPGe) and miles-per-charge. 
In practice, transit agencies trade off BEBs utilization and the reliability and cost-effectiveness of operating service; ultimately, they err conservatively and approach the ZEV transition anticipating the need for expanded fleets, oversized charging infrastructure, large onboard battery packs, and high electricity cost.
Reducing range and charging time uncertainty will reduce the above needs, and consequently the operating costs and capital expenditures of BEB fleets.

This paper presents a hierarchical approach to increasing BEB utilization in transit fleets using three key components.

\emph{ZEV Digital Twin}.
This module accurately predicts BEB charge consumption on a per-vehicle, per-driver, and per-route basis, and accurately predicts BEB charging time on a per-vehicle and per-charger basis.
Learning is employed to build high-accuracy prediction models of BEB energy consumption and charging time, which are a fundamental building block of energy-aware algorithms~\cite{Guanetti2018,Sun2018,Sautermeister2018,Vatanparvar2018,Scheubner2019,Ayman2022,Wang2017,Tian2016,RINALDI2020102070,FIORI2021102978,9204749,9478303}.

\emph{Predictive Block Assignment}.
The learned digital twins accurately predict the BEBs charge consumption on each block and and time-to-charge their batteries.
These predictions are used to maximize BEB fleet utilization through predictive optimal block assignment, which computes the optimal assignment of blocks to BEBs and scheduling of BEB charging sessions.
In the recent literature, variations of this problem have been formulated, e.g. to minimize the overall energy consumption of the fleet \cite{Sivagnanam2021,Paul2014,Yang2020}, assuming the availability of battery swapping or fast charging \cite{Li2014}, decoupling the vehicle assignment and charge management problems \cite{Zhou2020,Olsen2020}; related but different problems have also been studied, such as the optimization of the fleet mix \cite{Santos2016}, optimization of the charging infrastructure \cite{Lotfi2020}, schedule optimization for electric buses \cite{Jefferies2020,Picarelli2020} vehicle routing problems with energy constraints and/or mixed fleets \cite{Sassi2015}.

\emph{Depot Parking and Charging Queue Management}.
We propose a scheme to park the vehicles based on their charging demands. 
This module pre-computes trajectories for BEBs to maneuver inside the depot efficiently, get to a charging dispenser in time, and charge according to the charge strategy assigned by the predictive block assignment.
If autonomous driving and autonomous charging are not available, the output of this module can be used as a reference for human drivers and operators.

The implementation of the proposed approach coupled with funding support by federal, state, and local governments can accelerate the transition of the US transit fleet to ZEVs; which in turn can provide massive reductions of GHG emissions providing cleaner air quality and less noise pollution to communities.

The paper is structured as follows: in Section~\ref{sec:transit} we provide some background on ZEV transit fleets and on electric mobility on demand services; Section~\ref{sec:architecture} describes the proposed architecture; Section~\ref{sec:learning} the ZEV Digital Twin module; Section~\ref{sec:assignment} describes the Predictive Block Assignment module; Section~\ref{sec:parking} discusses the Depot Parking and Charging Queue Management.
The paper concludes with an example in Section~\ref{sec:example} which highlights the benefits of the proposed approach compared to the current approach.

\section{ZEV fleets in fixed schedule transit}
\label{sec:transit}

In this section, we provide a short background on transit fleets and the challenges of introducing  BEBs.
In general, transit agencies strive to avoid BEB stranding due to lack of charge (which causes service interruption and the need to send a replacement bus and to tow the standed bus), reducing the cost of service (due to nonlinear increases of electricity cost asSOCiated with high power and peak hour demand), and maximizing BEB usage and longevity.
Next, we describe some of the transit operations for which the introduction of BEBs is most challenging.

\subsection{Block Assignment}

In the context of transit agencies, a (fixed) schedule defines the routes, each visiting a set of stop locations, that will be served each day of the week for a certain period of time (e.g a few months).
In this context, a \emph{service trip} refers to a route driven at a certain time of the day and day of the week, so that the corresponding bus stop locations are visited at  pre-defined arrival and departure times.
A good schedule identifies the stops and trips necessary to satisfy public demand, maximizes service, minimizes operator cost, and meets the fleet size and operator headcount.
Trips are bundled into \emph{blocks}, units of work to be performed by a certain bus and driver.
Blocks generally start and end in one or a few locations (parking lots and depots with maintenance, fueling, and charging infrastructure).
A good block minimizes the non-service miles that need to be driven between trips that start and end at different locations (also known as \emph{deadhead trips}); at the same time, a block should normally be feasible for the driving range of a single bus.

Vehicle assignment is the problem of assigning blocks to buses on a daily basis, depending on their availability.
In traditional bus operations, buses of the same type (e.g. occupancy and geometry) are interchangeable between blocks.
In contrast, BEBs assignment to certain blocks must be considered carefully, due to the limited driving range which depends on battery capacity, current SOC level, and the varying energy demand of each BEB on different blocks.
Currently, transit agencies cannot accurately predict the feasibility of a block for a certain BEB, nor its SOC when it will return to the depot, nor how long it will take to charge it to a certain SOC such that it can run another block later in the day.
Depot operators make such predictions based on simplistic models such as MPGe and miles-per-charge, and on empirical rules on which blocks can or should be assigned to BEBs.
These models only capture the block distance and neglect other critical factors for energy consumption (such as traffic, weather, topography, driver behavior, and passenger occupancy); the resulting predictions have high uncertainty, which in turn leads to over-conservatism and poor utilization of BEBs.
In this paper, we assume that both the schedule and the bundling of trips into blocks are given, and focus on automating the assignment of BEBs to blocks.

\subsection{Off-route or depot charging}

Off-route or depot charging refers to charging the BEB battery outside of its service trips, typically at a depot equipped with charging infrastructure.
BEBs may be charged both overnight (after their daily assignments have been completed) and during the day (between different blocks).
Often, vehicles are assigned to blocks only up to 1-2 hours before pull-out; thus, even if the energy required by each BEB on each possible block is predicted with high accuracy, the decision to charge a BEB has to be made, in most cases, before the next block is known.
In other words, scheduling depot charging sessions is tightly coupled with the assignment of vehicles to blocks.
The current practice utilizes the depot chargers suboptimally; a simple approach is to charge BEBs as soon as they get back to the depot and until they reach a sufficiently high SOC.
Finally, a certain amount of ``opportunistic'' charging is desirable, as it potentially allows BEBs to be used as relief vehicles, and not just for their assigned blocks. 

Variable utility rates and charging infrastructure costs provide a set of incentives and constraints.
It is desirable to maximize the amount of charging during off-peak hours when the cost per kWh is lower, while completing the charge prior to the next pull-out time; BEBs plugged in at the end of a shift may cost more to charge than if they delayed charging until the off-peak hours.
It is also desirable to minimize the peak charging power used, as the utility rate also depends on the maximum power needed from the grid. 
Charging power is a nonlinear function of the vehicle SOC: it is normally highest in the central SOC range (between 20\% and 60\%-80\%) and lowest at the SOC bounds. 


Finally, we note that it is more cost-effective to operate on a lower ``charger to bus'' ratio (i.e., not having a charger dedicated to each BEB); this, however, makes it impossible to simultaneously charge all the BEBs in the fleet, and enhances the need for accurate modeling of energy demand and of charging time, as well as of depot charging queues. 

\section{System Architecture}
\label{sec:architecture}

\begin{figure}[ht]
\begin{center}
\includegraphics[width=\linewidth]{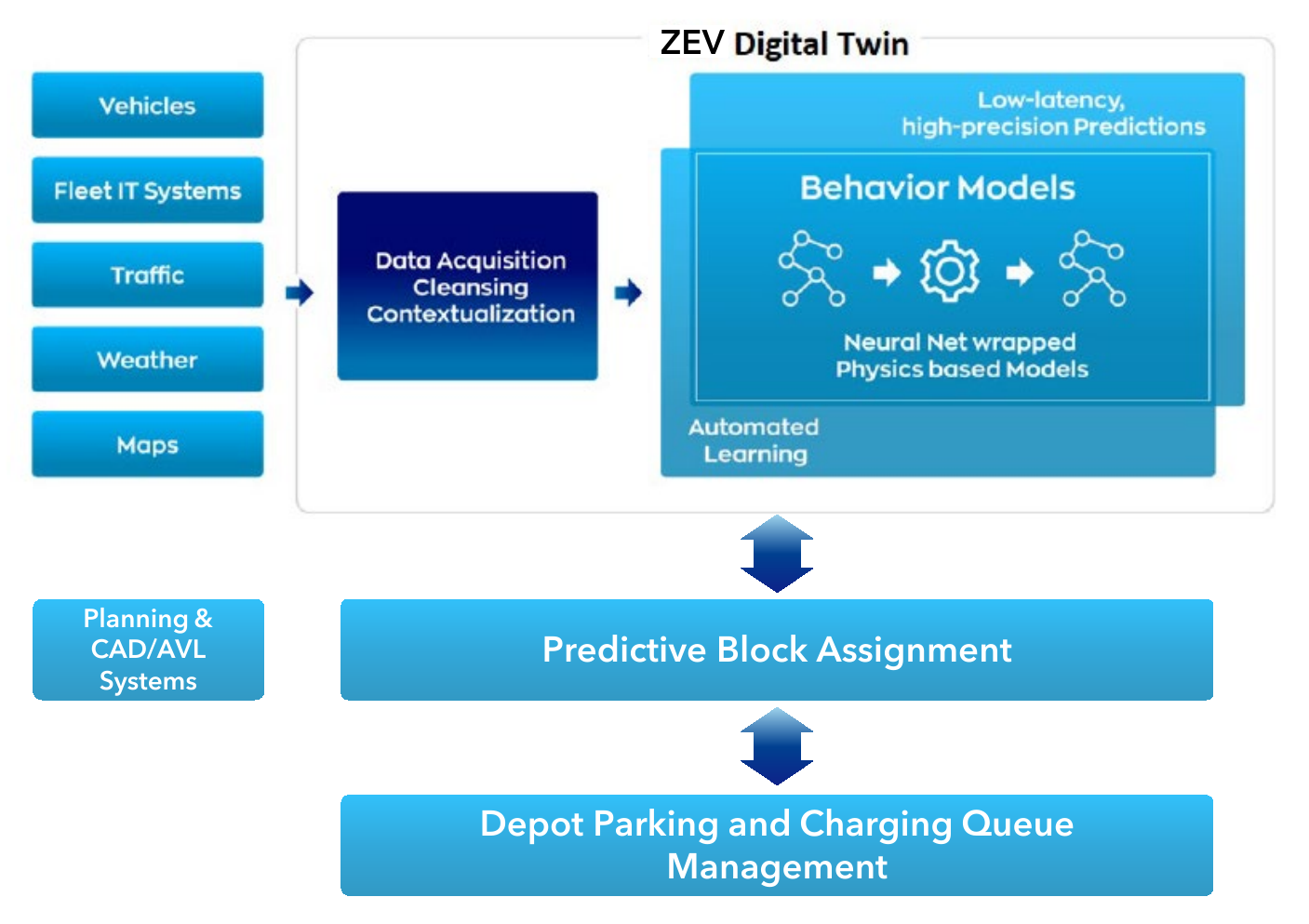}    
\caption{{Automated ZEV fleet operation architecture.}}
\label{fig0}
\end{center}
\end{figure}

Fig.~\ref{fig0} shows the proposed architecture for automated ZEV fleet operation.
The first component is an IoT platform which automatically learns ZEV digital twins based on physics principles (vehicle motion, components efficiency) and data-driven approaches (human factors, control policies, uncertainty modeling). 
The key aspects of this platform are scalability, ease of deployment, and the ability to operate reliably with minimal expert supervision. 
The resulting models predict vehicle performance on a per-vehicle, per-driver, per-mission basis, and feed model-based control and decision-making algorithms with predictions of provable accuracy.

The second component is an automated and optimal block assignment and charging scheduling module.
The goal of the optimization problem is to assign blocks to buses  as well as decide when and for how long to charge. 
The optimal control problem uses the digital twin model and the state of charge of each BEB in order to ensure that the resulting block assignment and charging schedule are feasible from an energy and time point of view. 
The cost function is formulated so as to maximize the total BEB utilization.

The last component is a parking and charging module which manages the motion of the BEBs inside the depot and pre-computes trajectories for BEBs to maneuver inside the depot efficiently, get to a charging dispenser in time, and charge according to the optimal charging schedule.

Note that the bidirectional arrows are used to represent feedback mechanisms between the modules of the architecture. 
For instance, if a BEB cannot be charged to the desired level in time, the low-level module will ask the \emph{Predictive Block Assignment} module for a new assignment based on the current unplanned state.


\section{Automated Learning of ZEV Digital Twins}
\label{sec:learning}


Obtaining high-precision predictive models of vehicle energy consumption is challenging as energy consumption is greatly affected by a long list of factors that depend on the specific vehicle, driver, and driving environment. 
These factors include speed profile, vehicle load, road gradient and curvature, ambient temperature, wind speed, state, and aging of the vehicle components. 



We have  developed and deployed a platform which delivers high-precision energy consumption prediction for EVs and is tailored for real-time optimization.
The platform leverages real-time vehicle telematics data streams, as well as third party maps, traffic and weather forecasts, and delivers tailored route and charging recommendations.
The platform consists of (i) models capturing the energy consumption and charging time on a per-vehicle, per-driver, per-road-segment, per-charger basis, and (ii) data-driven learning algorithms that estimate from data the model parameters to deliver high precision charge consumption estimation and charge depletion trajectories on routes for electrified vehicles. 
The learned models combine physics based and data driven modeling principles and are amenable for automated learning from data over time, such as to adjust to degradation of vehicle components, including the drivetrain, battery, and motor. 

As an example, the physics-based model below can describe the traction power demand of a two-wheel drive EV:
\begin{align*}
    P_{t,k} &= \frac{F_{m,k} v_k}{\eta_m(F_{m,k}, v_k, \theta_{m,k})}, \\
    F_{m,k} &= F_{k} - \psi_b(F_k, v_k, a_k), \\
    F_k &= M_{\tau} g \left( \sin \alpha(s_k) + C_{r}(s_k, v_k, p_k) \cos \alpha(s_k) \right) \\
    &+ C_{a}(\sigma_{\tau}, v_k, v_{w,k}, \phi_k, \phi_{w,k}) v_k^2 \\
    &+ \tilde{M}_{\tau} a_k + \frac{M_{\tau}^2}{4 C_s \rho^2(s_k)} v_k^3,
\end{align*}
where $k$ is a time index, $\tau$ is a trip or route leg index, $F_m$ and $F_b$ are the forces from electric motor and friction brakes (defined at the wheel), $P_t$ is the electrical motor power used for traction, $v$, $a$ and $\phi$ are the vehicle speed, acceleration, and bearing, $s$ is a curvilinear abscissa along the vehicle's path, $\alpha$ and $\rho$ are the road gradient and curvature radius, $M$ is the mass of the vehicle and its occupants, $\tilde{M}$ is $M$ plus the rolling inertias, $v_w$ and $\phi_w$ is the wind speed and bearing, $C_s$ is the wheel cornering stiffness.
$C_r$ is a rolling resistance coefficient, defined as a nonlinear function of the road segment, vehicle speed, and tire pressure $p$.
$C_a$ is an aerodynamic coefficient, defined as a nonlinear function of the vehicle and the wind's speed and bearing; $ \sigma$ takes discrete values corresponding to different configurations of the vehicle and its trailers and roof boxes.
$\psi_b$ is the function that allocates braking force to the hydraulic brakes, defined as a nonlinear function of speed and acceleration.
$\eta_m$ is the efficiency map of the electric motor, defined as a nonlinear function of the motor torque, speed, and temperature. 
All the above parameters and mappings are learned from data.
Similar models are derived for the auxiliaries, the battery, the driver, and are here omitted for brevity.

\begin{figure}
\begin{center}
\includegraphics[width=0.75\linewidth]{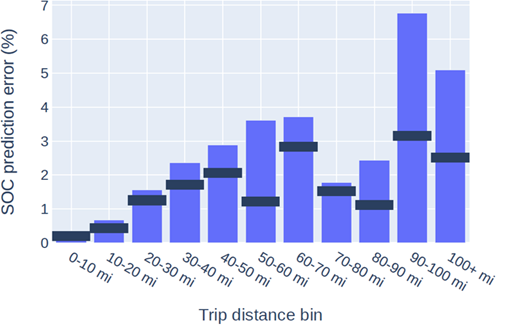}    
\caption{EV charge consumption prediction error (average and worst case) using the proposed approach.}
\label{fig3}
\end{center}
\end{figure}

Accurate prediction of the trip charge consumption has been demonstrated to be greater than 90\% in 87\% of trips on 50,000 miles of road trials (see Fig.~\ref{fig3}).

\section{ZEV optimal block assignment and charging strategy for fixed schedule transit}
\label{sec:assignment}

This section presents the optimal block assignment and charging strategy for a mixed ZEV fleet, enabled by accurate predictions of block charge consumption and battery time-to-charge. 
The objectives are to optimize the assignment of vehicles to blocks for maximum BEB utilization, and the charging schedule across a transit fleet, defining when and how much each BEB should be charged, while ensuring that BEBs are safely operated within their minimum and maximum SOC, that physical constraints (such as the number and displacement of available electric chargers) and bus-to-block constraints (certain blocks can only be completed by certain bus types, for instance, due to the bus size and the route geometry) are respected. 

First, we introduce some notation and definitions. 
Let $B$ be the set of available buses and each bus $b\in B$ belong to a certain bus type $b_t \in B_t$ capturing the manufacturer, model, trim, powertrain (internal combustion, hybrid electric, battery electric, fuel cell), capacity of energy storage (such as fuel tank capacity or battery capacity), rated driving range, maximum occupancy, weight, and payload.
We also asSOCiate each bus to a digital twin model $b_m \in B_m$ that can predict energy usage and charging time.
The buses in $B$ serve a set of blocks $J$.
Each block $j \in J$ includes both service trips and deadhead trips, and has an origin and destination (which in the single depot case is the same), and a start and end time.
An assignment plan defines, on a given day, the blocks that each bus will serve. 

Since we are focusing on fleets that include some BEBs, a feasible assignment plan must ensure that BEBs are only assigned to blocks that can be completed without getting stranded.
Such feasibility depends on the SOC at the beginning and at the end of the day, on the minimum and maximum SOC allowed for the BEB, on the charge consumption incurred on the assigned blocks, and on the feasibility of charging the BEB up to the SOC required for the next block while it is at the depot.
Similar to the assignment plan, the charging strategy defines, on a given day, the charging sessions for each BEB; each charging session is asSOCiated with a charging location in the depot, and a start and end time. 


We can now formulate the fixed time optimization problem for the optimal block assignment and charging strategy for the BEBs in a fixed schedule transit fleet.
The decision variables for the optimization problem include the binary variables $b_{i,j}$ indicating whether a bus $i \in B$ is serving a block $j\in J$, and the binary variables $c_{i}(t)$ indicating whether a bus $i \in B$ is charging at time $t$. The optimization problem has the form:
\begin{subequations}
\label{eq:zev_assignment}
\begin{align}
\min_{b_{i,j},c_{i}(:)} \ & \ J = \displaystyle -\smashoperator{\sum_{i\in B}}\,\smashoperator{\sum_{ j \in J}} b_{i,j}d_{j} \\
\text{s.t. } \ & \textrm{SOC dynamics with $b_m$, } \\
& \textrm{vehicle initial and final state constraints, }\\
& \textrm{vehicle safety constraints, }\\
& \textrm{depot charging constraints, }\\
& \textrm{bus-to-block constraints, }
\end{align}
\end{subequations}
where $d_j$ is the distance (utilization) of block $j \in J$, 
and the SOC dynamics represent, for each bus $b \in B$, the predictions by model $b_m$ of the energy used to service a block and added in a charging session. 
The vehicle initial and final state constraints enforce initial and final states, such as battery SOC and location, for each BEB. 
The vehicle safety constraints ensure that the buses operate within the safe SOC levels.
The depot charging constraints model constraints imposed by the charging infrastructure in the depot; these can include the maximum number of parallel or serial charging, bus queue orders, the maximum number of buses being charged at a time, as well as (soft or hard) constraints dictated by utility rates and the need to limit peak power demand.
Finally, the bus-to-block constraints ensure that some blocks are only served by certain bus types $b_t$.

\section{Depot Parking and Charging}
\label{sec:parking}

This section presents the module to manage the motion of ZEVs inside the depot according to the charging schedule. The vehicles can be notified through dashboard messages or phone app, or autonomously controlled if equipped with a level 4 self-driving stack.

\subsection{Parking and Charging Management}

\begin{figure}
\begin{center}
\includegraphics[width=0.75\linewidth]{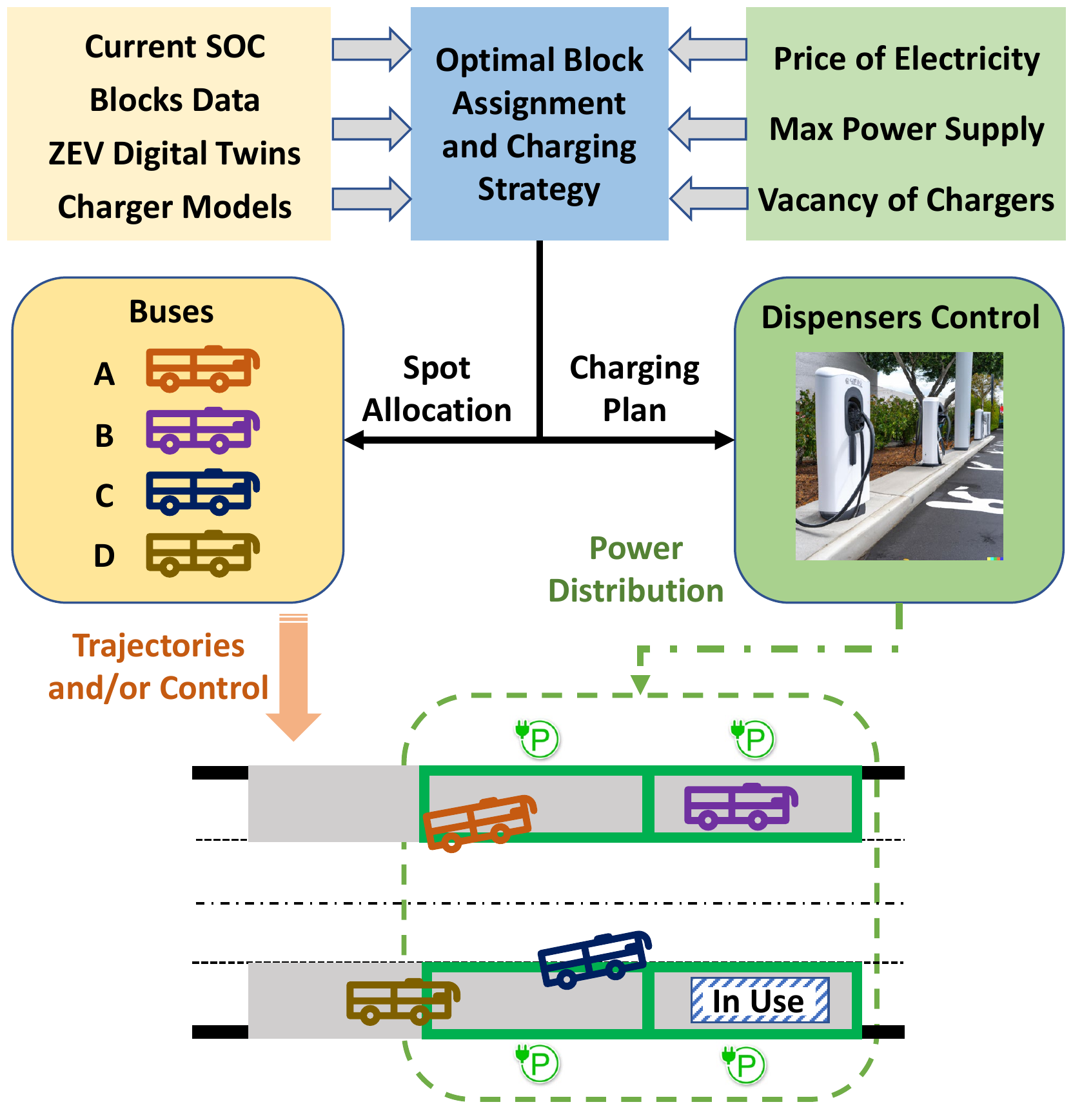}    
\caption{The system diagram of the parking and charging management module}
\label{fig:parking-charging}
\end{center}
\end{figure}

As illustrated in Fig.~\ref{fig:parking-charging}, when a fleet of ZEV enters the depot, they can report to the system the necessary parameters, such as their current state of charge (SOC), blocks data, ZEV digital twins, and charger models. By observing and predicting the parking and charging demand, the cloud server can solve the optimal block assignment and charging strategy~\eqref{eq:zev_assignment} and notify the vehicles of their assigned time and location of the parking spot, their planned charging schedule, and the path to reach the spots.

A ZEV can be either assigned immediately to spots with chargers (e.g., buses A, B, and C in Fig.~\ref{fig:parking-charging}) or tentatively assigned to parking-only spots on a waitlist (e.g., vehicle D). For the latter case, the system can coordinate the interchange of vehicles on a charger. When one of the buses finishes charging (e.g., vehicle C), it can be asked to leave the charging spot as soon as possible to serve a block. When it leaves, the vehicle on a waitlist (e.g., vehicle D) can be notified to move in and use the now-open charger.

\subsection{Autonomous Control of Vehicles}

If any of the ZEVs is autonomous with level 4 self-driving stack, it can self-drive along a designated trajectory to maneuver in and out of the charging spot~\cite{wang2014automatic}.

Given the initial state $z_{0}$, target state $z_{\mathrm{F}}$, and vehicle dynamics $\dot{z} = f(z, u)$ of a ZEV, we can solve the trajectory planning problem with the formulation:
\begin{subequations}
\label{eq:traj-planning}
\begin{align}
    \min_{\mathbf{z}, \mathbf{u}, T} \ & \ J = \int_{t=0}^{T} c\left(z(t), u(t)\right) dt\nonumber \\
    \text{s.t. } \ & \dot{z}(t) = f(z(t), u(t)), \\
    & z(t) \in \mathcal{Z}, u(t) \in \mathcal{U},\\
    & z(0) = z_0, z(T) = z_{\mathrm{F}}, \\
    &  \mathrm{dist}\left(\mathbb{B}(z(t)), \mathbb{O}^{[m]}\right) \geq d_{\mathrm{min}}, \forall m \label{eq:CA-constr}
\end{align}
\end{subequations}
where the state $z$ and input $u$ are constrained under operation limits $\mathcal{Z}$ and $\mathcal{U}$. We denote by $\mathbb{B}(z(t))$ the vehicle body at time $t$ and ask it to maintain a safety distance $d_{\mathrm{min}}$ away from all obstacles $\mathbb{O}^{[m]}$. The stage cost $c(\cdot, \cdot)$ can encode the amount of actuation, energy consumption, and time consumption.

\begin{figure}
\begin{center}
\includegraphics[width=0.75\linewidth]{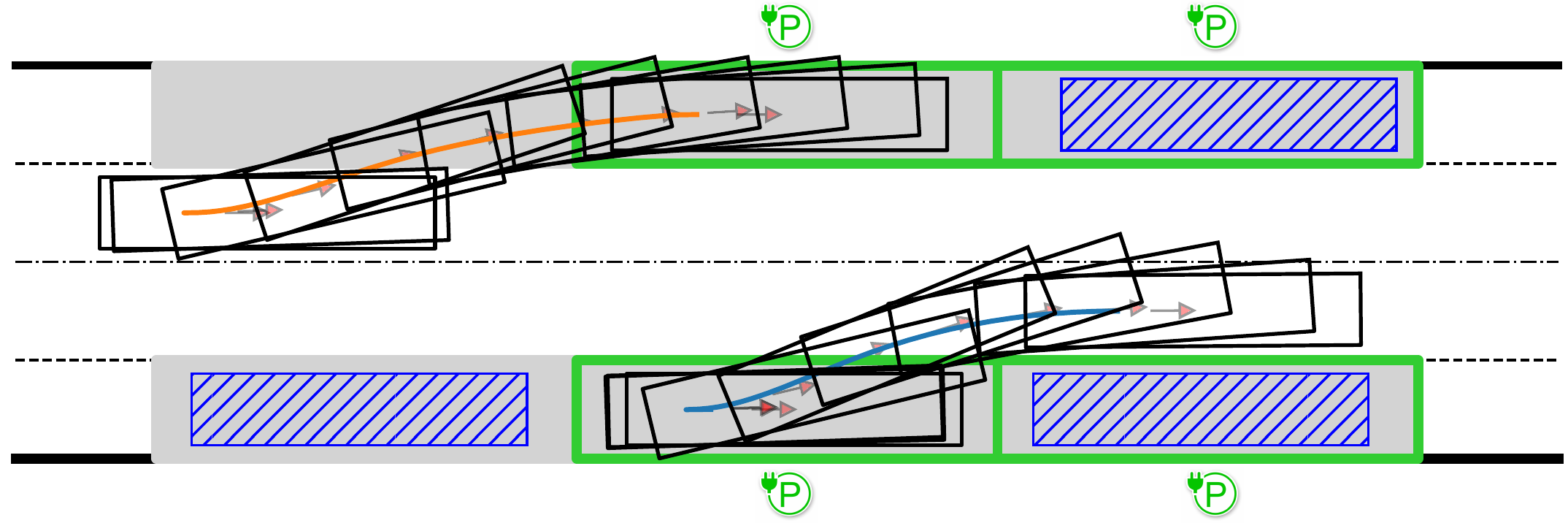}    
\caption{ZEV maneuver in tightly constrained depot}
\label{fig:maneuver}
\end{center}
\end{figure}

Problem~\eqref{eq:traj-planning} can be solved efficiently by reformulating~\cite{zhang_optimization-based_2020} the collision avoidance constraints~\eqref{eq:CA-constr} and discretizing with orthogonal collocation on finite elements~\cite{biegler_nonlinear_2010}. Fig.~\ref{fig:maneuver} shows the kinematically feasible, collision-free trajectories for buses A and C in Fig.~\ref{fig:parking-charging} to enter and leave their charging spots. Note that since the structure of the depot is known, and the transit ZEVs will always drive along dedicated lanes repetitively, we can enumerate all possible maneuvers and generate the trajectories in advance as a library~\cite{shen_fleet_2019}, and send them to buses according to the specific real-time scenario.

\section{Simulation results and discussion}
\label{sec:example}
In this section, we present simulation results of the ZEV block assignment and charging strategy \eqref{eq:zev_assignment}. 

We used samples of block data and bus block and charging models obtained from a US transit agency. 
A total of 8 BEBs are initially at 85\% SOC and ready for pull out at 5 am EST, and have 8 electric chargers available at the depot. 
The minimum SOC is set to 35\% throughout the plan for each BEB. 
To favor BEB utilization around the time of the day with the most active blocks (4 pm EST), an additional constraint is imposed to have at least 4 BEBs serving blocks at 4 pm EST. 
As for the depot charging constraints, in addition to the physical constraints due to the chargers power and displacement in the depot, it is imposed that there are at most 3 charging sessions active at the same time during peak hours (6 am and 10 pm EST), so that peak power demand and the corresponding utility rates are limited. 
We also allocate at least 15 minutes of setup time before any charging or service block begins.
Finally, every bus ends the day at 85\% SOC or above to be ready for service the next day.

\begin{figure}
\begin{center}
\includegraphics[width=0.9\linewidth]{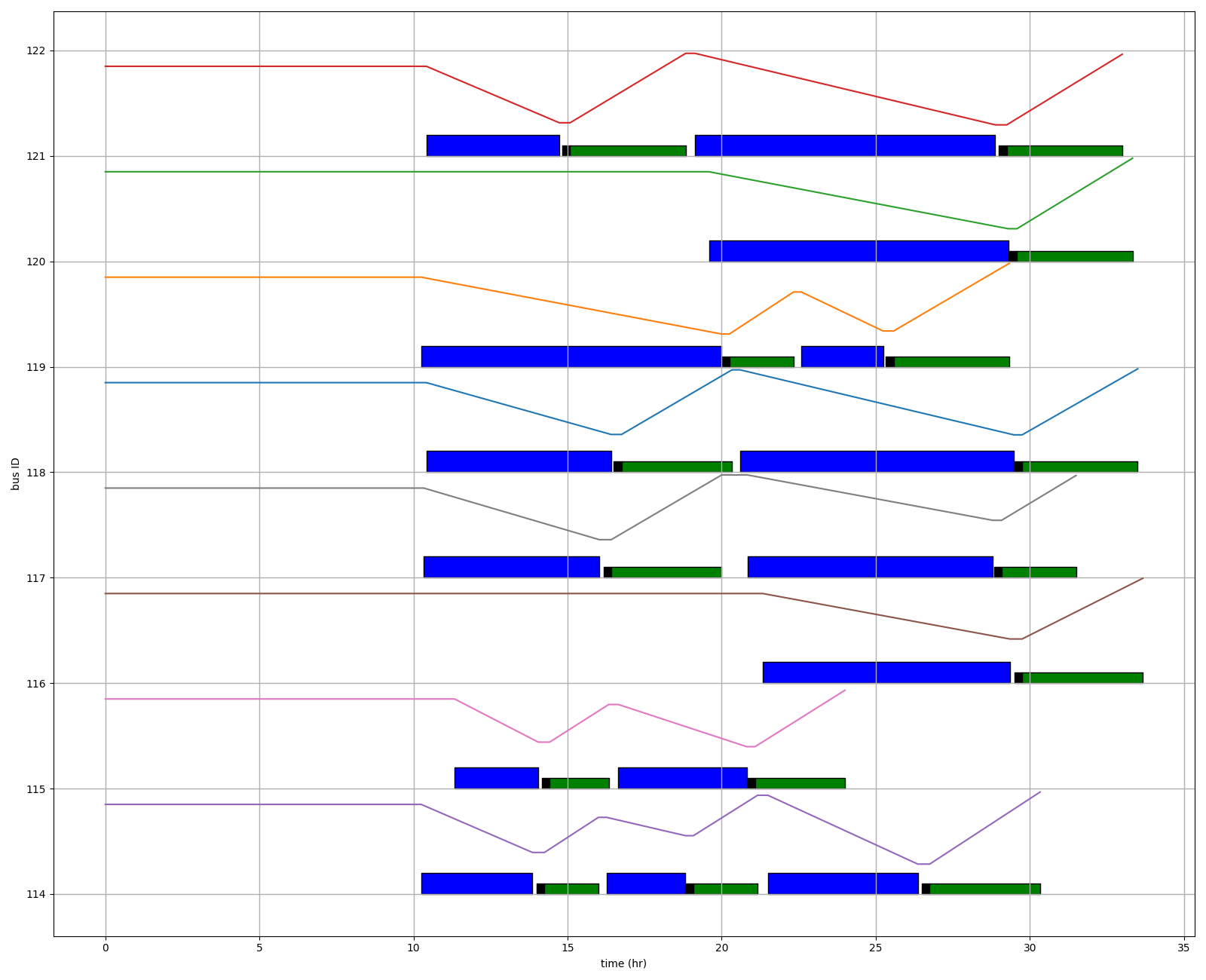}    
\caption{Block and charging assignments with perfect prediction; x- and y-axis are UTC time and bus IDs.}
\label{fig:perfect}
\end{center}
\end{figure}

\begin{figure}
\begin{center}
\includegraphics[width=0.9\linewidth]{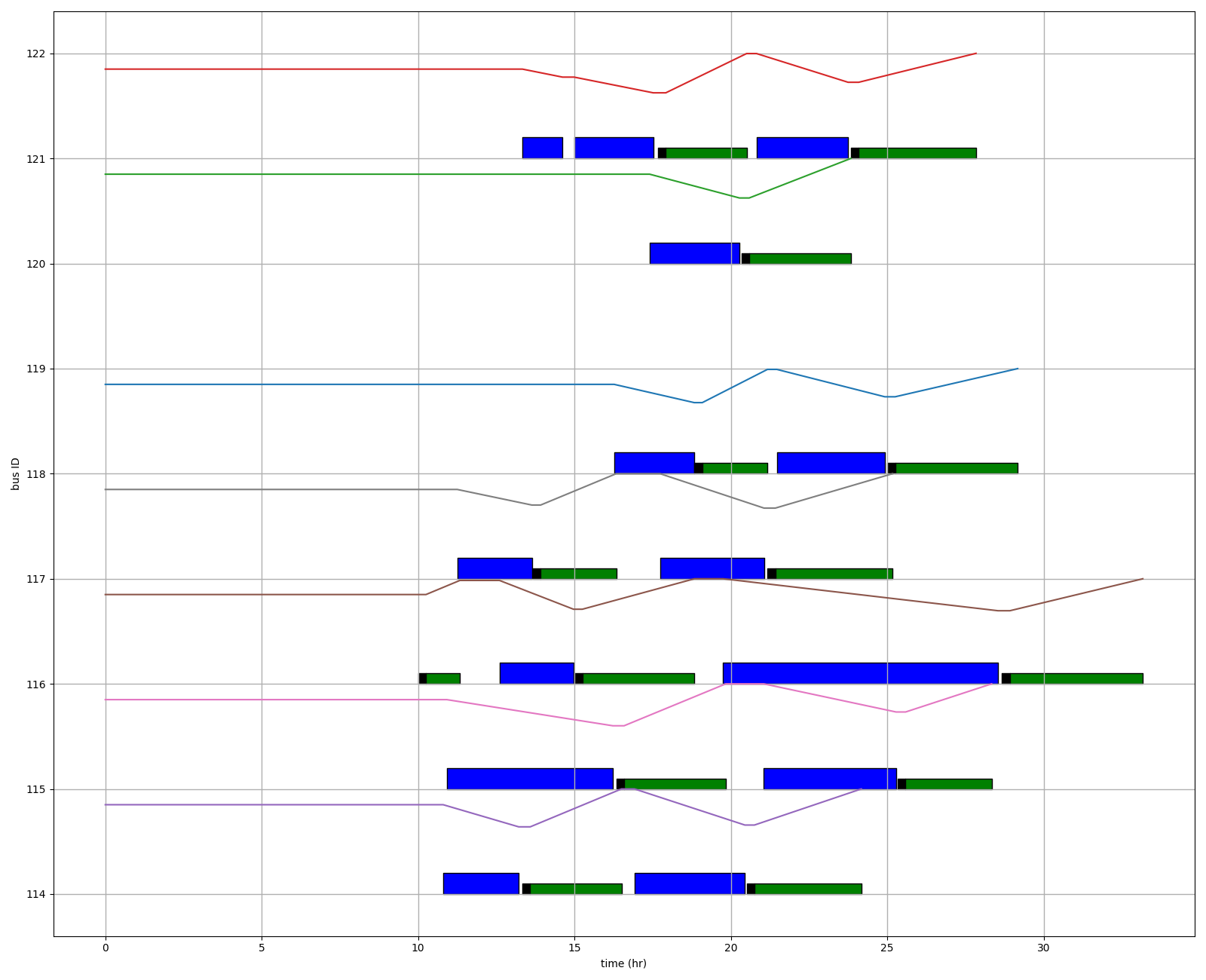}    
\caption{Block and charging assignments with imperfect predictions; x- and y-axis are UTC time and bus IDs. Actual block energy consumption is half of predicted consumption.}
\label{fig:imperfect}
\end{center}
\end{figure}

Fig.~\ref{fig:perfect} and Fig.~\ref{fig:imperfect} show the simulation results of the daily schedules and the simulated SOC trajectories of each bus. Fig.~\ref{fig:perfect} is the result when the SOC consumption of the service blocks and charging are the same as the SOC dynamics predicted by $b_m$, and Fig.~\ref{fig:imperfect} is the result when the actual SOC consumption of the service blocks is only half the predicted while the charging is accurate. 
Each row corresponds to a BEB, the blue rectangles represent the blocks served by each BEB, the green rectangles indicate that the bus is charging, and the black rectangles indicate the preparation time prior to charging.
The colored solid lines represent the corresponding SOC trajectories.

In both scenarios, every constraint stated in the problem details is satisfied, including the minimum SOC above 35\%, the final SOC above 85\%, the maximum number of charging session active during peak hours under 3, etc.
With accurate block energy consumption predictions, the assignment plan includes more and longer blocks, and the total distance traveled is 1648.4 miles. 
When the block energy consumption is overestimated, the assignment plan includes fewer smaller blocks, covering a total distance of 747.2 miles (less than half of the perfect prediction scenario).
It is also seen that one of the buses does not serve any blocks.


\bibliographystyle{IEEEtran}
\bibliography{references}

\begin{thebibliography}{10}
\providecommand{\url}[1]{#1}
\csname url@rmstyle\endcsname
\providecommand{\newblock}{\relax}
\providecommand{\bibinfo}[2]{#2}
\providecommand\BIBentrySTDinterwordspacing{\spaceskip=0pt\relax}
\providecommand\BIBentryALTinterwordstretchfactor{4}
\providecommand\BIBentryALTinterwordspacing{\spaceskip=\fontdimen2\font plus
\BIBentryALTinterwordstretchfactor\fontdimen3\font minus
  \fontdimen4\font\relax}
\providecommand\BIBforeignlanguage[2]{{%
\expandafter\ifx\csname l@#1\endcsname\relax
\typeout{** WARNING: IEEEtran.bst: No hyphenation pattern has been}%
\typeout{** loaded for the language `#1'. Using the pattern for}%
\typeout{** the default language instead.}%
\else
\language=\csname l@#1\endcsname
\fi
#2}}

\bibitem{Guanetti2018}
J.~Guanetti, Y.~Kim, and F.~Borrelli, ``{Control of Connected and Automated
  Vehicles: State of the Art and Future Challenges},'' \emph{Annual Reviews in
  Control}, vol.~45, pp. 18--40, 2018.

\bibitem{Sun2018}
C.~Sun, J.~Guanetti, F.~Borrelli, and S.~Moura, ``Optimal eco-driving control
  of connected and autonomous vehicles through signalized intersections,''
  \emph{IEEE Internet Things J.}, vol.~7, no.~5, pp. 3759--3773, 2020.

\bibitem{Sautermeister2018}
S.~Sautermeister, M.~Falk, B.~Baker, F.~Gauterin, and M.~Vaillant, ``Influence
  of measurement and prediction uncertainties on range estimation for electric
  vehicles,'' \emph{IEEE Trans. Intell. Transp. Syst.}, vol.~19, pp.
  2615--2626, 2018.

\bibitem{Vatanparvar2018}
K.~Vatanparvar, S.~Faezi, I.~Burago, M.~Levorato, and M.~A.~A. Faruque,
  ``Extended range electric vehicle with driving behavior estimation in energy
  management,'' \emph{IEEE Trans. Smart Grid}, vol.~14, pp. 1--10, 2018.

\bibitem{Scheubner2019}
S.~Scheubner, A.~T. Thorgeirsson, M.~Vaillant, and F.~Gauterin, ``A stochastic
  range estimation algorithm for electric vehicles using traffic phase
  classification,'' \emph{IEEE Trans. Veh. Technol.}, vol.~68, no.~7, pp.
  6414--6428, 2019.

\bibitem{Ayman2022}
A.~Ayman, A.~Sivagnanam, M.~Wilbur, P.~Pugliese, A.~Dubey, and A.~Laszka,
  ``Data-driven prediction and optimization of energy use for transit fleets of
  electric and ice vehicles,'' \emph{ACM Transactions on Internet Technology},
  vol.~22, 2022.

\bibitem{Wang2017}
Y.~Wang, D.~Zhang, L.~Hu, Y.~Yang, and L.~H. Lee, ``A data-driven and optimal
  bus scheduling model with time-dependent traffic and demand,'' \emph{IEEE
  Trans. Intell. Transp. Syst.}, vol.~18, pp. 2443--2452, 9 2017.

\bibitem{Tian2016}
Z.~Tian, T.~Jung, Y.~Wang, F.~Zhang, L.~Tu, C.~Xu, C.~Tian, and X.~Y. Li,
  ``Real-time charging station recommendation system for electric-vehicle
  taxis,'' \emph{IEEE Trans. Intell. Transp. Syst.}, vol.~17, pp. 3098--3109,
  11 2016.

\bibitem{RINALDI2020102070}
M.~Rinaldi, E.~Picarelli, A.~D'Ariano, and F.~Viti, ``Mixed-fleet
  single-terminal bus scheduling problem: Modelling, solution scheme and
  potential applications,'' \emph{Omega}, vol.~96, p. 102070, 2020.

\bibitem{FIORI2021102978}
C.~Fiori, M.~Montanino, S.~Nielsen, M.~Seredynski, and F.~Viti, ``Microscopic
  energy consumption modelling of electric buses: model development,
  calibration, and validation,'' \emph{Transportation Research Part D:
  Transport and Environment}, vol.~98, p. 102978, 2021.

\bibitem{9204749}
S.~Hulagu and H.~B. Celikoglu, ``An electric vehicle routing problem with
  intermediate nodes for shuttle fleets,'' \emph{IEEE Trans. Intell. Transp.
  Syst.}, vol.~23, no.~2, pp. 1223--1235, 2022.

\bibitem{9478303}
------, ``Electric vehicle location routing problem with vehicle motion
  dynamics-based energy consumption and recovery,'' \emph{IEEE Trans. Intell.
  Transp. Syst.}, vol.~23, no.~8, pp. 10\,275--10\,286, 2022.

\bibitem{Sivagnanam2021}
\BIBentryALTinterwordspacing
A.~Sivagnanam, A.~Ayman, M.~Wilbur, P.~Pugliese, A.~Dubey, and A.~Laszka,
  ``Minimizing energy use of mixed-fleet public transit for fixed-route
  service,'' 2021. [Online]. Available: \url{http://arxiv.org/abs/2004.05146}
\BIBentrySTDinterwordspacing

\bibitem{Paul2014}
T.~Paul and H.~Yamada, ``Operation and charging scheduling of electric buses in
  a city bus route network.''\hskip 1em plus 0.5em minus 0.4em\relax Institute
  of Electrical and Electronics Engineers Inc., 11 2014, pp. 2780--2786.

\bibitem{Yang2020}
X.~Yang and L.~Liu, ``A multi-objective bus rapid transit energy saving
  dispatching optimization considering multiple types of vehicles,'' \emph{IEEE
  Access}, vol.~8, pp. 79\,459--79\,471, 2020.

\bibitem{Li2014}
J.-Q. Li, ``Transit bus scheduling with limited energy,'' \emph{Transportation
  Science}, vol.~48, pp. 521--539, 2014.

\bibitem{Zhou2020}
G.~Zhou, D.~Xie, X.~Zhao, and C.~Lu, ``Collaborative optimization of vehicle
  and charging scheduling for a bus fleet mixed with electric and traditional
  buses,'' \emph{IEEE Access}, vol.~8, pp. 8056--8072, 2020.

\bibitem{Olsen2020}
N.~Olsen, N.~Kliewer, and L.~Wolbeck, ``A study on flow decomposition methods
  for scheduling of electric buses in public transport based on aggregated
  time–space network models,'' \emph{Central European Journal of Operations
  Research}, 2020.

\bibitem{Santos2016}
D.~Santos, Z.~Kokkinogenis, J.~F.~D. Sousa, D.~Perrotta, and R.~J. Rossetti,
  ``Towards the integration of electric buses in conventional bus
  fleets.''\hskip 1em plus 0.5em minus 0.4em\relax Institute of Electrical and
  Electronics Engineers Inc., 12 2016, pp. 88--93.

\bibitem{Lotfi2020}
M.~Lotfi, P.~Pereira, N.~Paterakis, H.~A. Gabbar, and J.~P. Catalao,
  ``Optimizing charging infrastructures of electric bus routes to minimize
  total ownership cost,'' vol. 029803, 2020.

\bibitem{Jefferies2020}
D.~Jefferies and D.~Göhlich, ``A comprehensive tco evaluation method for
  electric bus systems based on discrete-event simulation including bus
  scheduling and charging infrastructure optimisation,'' \emph{World Electric
  Vehicle Journal}, vol.~11, 2020.

\bibitem{Picarelli2020}
E.~Picarelli, M.~Rinaldi, A.~D'Ariano, and F.~Viti, ``Model and solution
  methods for the mixed-fleet multi-terminal bus scheduling problem,''
  \emph{Transportation Research Procedia}, vol.~47, pp. 275--282, 2020.

\bibitem{Sassi2015}
O.~Sassi, W.~R. Cherif, and A.~Oulamara, ``Vehicle routing problem with mixed
  fleet of conventional and heterogenous electric vehicles and time dependent
  charging costs,'' \emph{Int. J. of Math., Comp., Phys., Electrical and
  Computer Engineering}, vol.~9, pp. 167--177, 2015.

\bibitem{wang2014automatic}
W.~Wang, Y.~Song, J.~Zhang, and H.~Deng, ``Automatic parking of vehicles: A
  review of literatures,'' \emph{International Journal of Automotive
  Technology}, vol.~15, pp. 967--978, 2014.

\bibitem{zhang_optimization-based_2020}
X.~Zhang, A.~Liniger, and F.~Borrelli, ``Optimization-{Based} {Collision}
  {Avoidance},'' \emph{IEEE Trans. Control Syst. Technol.}, vol.~29, no.~3, pp.
  972--983, 2021.

\bibitem{biegler_nonlinear_2010}
L.~T. Biegler, \emph{Nonlinear {Programming}: {Concepts}, {Algorithms}, and
  {Applications} to {Chemical} {Processes}}.\hskip 1em plus 0.5em minus
  0.4em\relax USA: Society for Industrial and Applied Mathematics, 2010.

\bibitem{shen_fleet_2019}
X.~Shen, X.~Zhang, and F.~Borrelli, ``Autonomous parking of vehicle fleet in
  tight environments,'' in \emph{ACC 2020}, 2020, pp. 3035--3040.

\end{thebibliography}

\end{document}